\documentclass[%
superscriptaddress,
 amsmath,
 amssymb,
 aps,
 prl,
floatfix,
twocolumn
]{revtex4-2}
\usepackage[utf8]{inputenc}
\usepackage[T1]{fontenc}
\usepackage{babel}
\usepackage{graphicx}
\usepackage{amsfonts}
\usepackage{amssymb}
\usepackage{amsmath}
\usepackage{amsthm}
\usepackage{mathtools}
\usepackage{physics}
\usepackage[normalem]{ulem}
\usepackage{blindtext}
\usepackage{dsfont}

\usepackage{mathtools}
\usepackage{braket}

\usepackage{graphicx}
\usepackage[format=plain,justification=centerlast]{caption}
\usepackage[format=plain,justification=centerlast]{subcaption}
\usepackage{bm}
\usepackage[dvipsnames]{xcolor}

\usepackage{natbib}
\setcitestyle{square,numbers}
\usepackage[colorlinks]{hyperref}

\begin{document}

\title{No-go for fully unitary quantum mechanics from Bell's Theorem; comment on ``Physics and Metaphysics of Wigner’s Friends: Even performed pre-measurements
have no results''} 
\author{Konrad Schlichtholz}
\affiliation{International Centre for Theory of Quantum Technologies, University of Gdansk, 80-308 Gda{\'n}sk, Poland}
\email[Corresponding author: ]{konrad.schlichtholz@phdstud.ug.edu.pl}

\pagenumbering{arabic}

\begin{abstract}
    The purpose of this comment is to show that a reinterpretation of the results from the Letter: ``Physics and Metaphysics of Wigner’s Friends: Even performed pre-measurements
have no results'' allows for reaching the conclusion ``pre-measurements
have no results''  based only on postulates of quantum mechanics without additional assumptions on irreversibility. Additionally, with supplementary reasoning based on Bell's theorem, one can show that unitary decoherence cannot be solely responsible for the quantum-to-classical transition, and an additional irreversibility model is required for its full description. Consequently, the black hole information paradox has no physical basis. 
\end{abstract}

\maketitle

This comment presents a reinterpretation of the results from the Letter \cite{ZM}.
The authors of \cite{ZM} point out that, in a restatement of Wigner's Friend proposed in \cite{FR}, a paradox appears due to equating irreversible measurements with reversible pre-measurements to which one should not assign outcomes.   They argue that the inclusion of decoherence in Friend's system rules out the paradox. Based on this, they reject the statement from \cite{FR} ''quantum theory cannot consistently describe the use of itself''.

Let me start with the reanalysis of the GHZ-like reasoning (`Step three' in the Letter). In this thought experiment, based on Born's rule, `outcomes' of Friends' pre-measurements (unitary evolution of Friends' labs)  $f_i=\pm1$ and Wigners' measurements $w_i=\pm1$ are forced to obey relation $(f_1 f_2 f_3)^2 (w_1 w_2 w_3)^2=-1$. The authors argue that Friends cannot associate the outcomes $f_i$ as this leads to the paradox. However, this argument is based on the assumption that Wigners' outcomes are assigned correctly. The motivation for this is that Friends' outcomes can be wiped out by Wigners' actions, while Wigners' results cannot be undone due to effectively irreversible still unitary decoherence. Irreversibility arises since the control over degrees of freedom of macroscopic objects is effectively impossible.
This is not satisfactory, as this irreversibility is not intrinsic to the theory itself, and mathematical consistency is based on lack of contradiction between all theory's constituents --- not only between those practically implementable.
Therefore, at this stage, there is no particular mathematical reason for the irreversibility assumption, and this reasoning is just an improved version of the original paradox from \cite{FR}. However, one can modify the experiment assuming that Wigners also perform pre-measurements. In this case, the probabilities are assigned due to Born's rule in the same way, and thus the relation for the outcomes still holds. Therefore, even if only pre-measurements are done, we get the paradox. This shows without additional assumptions that one cannot associate outcomes with pre-measurements. However, associating outcomes with collapse does not cause paradoxes in such scenarios.

One can see decoherence emerging as a non-unitary Kraus representation of the evolution \cite{Deco} of the reduced density matrix of a subsystem (Friend) coupled with an environment (Friend's lab) where the whole system evolves unitarily. Therefore, such decoherence is already taken into account in this modified setup, as in fact it is pre-measurement. This can be seen from Eq. (2) in \cite{ZM}, as the reduced density matrix of Friend after unitary evolution of the entire lab is a classical mixture of states corresponding to possible results. At this point, one could try to assume that Friend assigned some outcome. However, as discussed, this causes paradox. Thus, unitary decoherence does not provide a consistent way of describing measurement, but rather it is its precursor, and therefore it does not fully describe the quantum-to-classical transition. 

This can also be seen in another way. In general, evolving to a classical mixture in a subsystem is not equivalent to assigning the result. Consider entanglement swapping \cite{Swapp} in which two singlets decoupled from the environment are recombined to form a new singlet. Treating the second singlet as part of the environment,  the reduced density matrix of modes of the first singlet after the procedure is simply the classical mixture.
However, we know that there is no local-hidden variable description of the system, and thus no meaningful outcome can be ascribed to it. This idea can be generalized to GHZ states generation \cite{GHZ,GHZ2} and scaled to arbitrary entangled system size. In general, unitary evolution entangles the system with the environment, whereas from Bell's theorem we know that the phenomenon of entanglement is governed without local-hidden variables (outcomes). Therefore, for a theory to describe unambiguously measurement on the subsystem, there has to be a moment when entanglement with the subsystem is completely removed on all levels, leading to all reduced density matrices containing the subsystem being block-diagonal with respect to measurement basis in this subsystem. This requires nonunitary behavior of the system at some scale of dilution of entanglement (e.g. discontinuous cut of coherences).  Therefore, observable multiparty Bell non-classicality provides indirect experimental proof that current unitary quantum mechanics does not describe measurement and is only an effective statistical approximation not fundamental to macroscopic systems. 
Simply, outcomes are not well defined in unitary evolution, and one needs to impose irreversibility to overcome that.

The above results show that information is irreversibly lost during evolution, and the black hole information paradox \cite{black} that builds upon the unitarity of evolution is not of fundamental concern.

\begin{acknowledgements}
Acknowledgements -- The author thanks M. Markiewicz for the discussion. The work is part of ‘International Centre for
Theory of Quantum Technologies’ project (contract no. 2018/MAB/5), which is carried out
within the International Research Agendas Programme (IRAP) of the Foundation for Polish
Science (FNP) co-financed by the European Union from the funds of the Smart Growth
Operational Programme, axis IV: Increasing the research potential (Measure 4.3).
\end{acknowledgements}
\bibliography{biblio}
\end{document}